\documentclass[aps, amssymb, amsmath, superscriptaddress, 
twocolumn] {revtex4-2}

\usepackage{chngcntr}

\usepackage{graphicx}
\usepackage{color}
\usepackage{amsmath}
\usepackage{enumitem}
\usepackage{amssymb}
\usepackage{hyperref}
\usepackage{cancel}
\usepackage{ulem}
\usepackage{multirow}

\newcommand{\be}{\begin{equation}}
\newcommand{\ee}{\end{equation}}

\newcommand{\bea}{\begin{eqnarray}}
\newcommand{\eea}{\end{eqnarray}}

\newcommand{\p}{\partial}

\newcommand{\la}{\left\langle}
\newcommand{\ra}{\right\rangle}
\newcommand{\lb}{\left[}
\newcommand{\rb}{\right]}
\newcommand{\lp}{\left(}
\newcommand{\rp}{\right)}

\newcommand{\tr}{{\rm \, tr\,}}

\renewcommand{\vec}[1]{{\boldsymbol #1}}

\newcommand{\addLL}[1]{\textcolor{magenta}{#1}}
\newcommand{\addQ}[1]{\textcolor{red}{#1}}

\makeatother

\begin{document}
	\title{Spin chirality and fermion stirring in topological bands} 
	
	\author{Archisman Panigrahi}
	\affiliation{Department of Physics, Massachusetts Institute of Technology, Cambridge MA02139, USA}
	\author{Vladislav Poliakov} 
\affiliation{Department of Physics, Massachusetts Institute of Technology, Cambridge MA02139, USA}
\author{Zhiyu Dong}
\affiliation{Department of Physics, California Institute of Technology, Pasadena, California 91125}
\author{Leonid Levitov}
\affiliation{Department of Physics, Massachusetts Institute of Technology, Cambridge MA02139, USA}

	\begin{abstract}
We demonstrate that in metals, both normal and superconducting, orbital currents present in the ground state when time reversal symmetry (TRS) is broken, generate spin chirality.
Nonzero chirality can 
emerge in the absence of any spin-dependent interactions, 
even when the ground state remains spin-unpolarized. 
The  chirality effect is derived diagrammatically and 
illustrated 
for Haldane model and the topological superconductivity problem. Chirality in the carrier band 
results in a chiral three-spin RKKY interaction between localized spins coupled to carriers by s-d Hamiltonian, an effect that can be detected by local probes such as spin-sensitive STM. In systems where detecting TRS breaking by conventional means is challenging, such as topological superconductors, local detection of spin chirality can serve as a reliable diagnostic of superconducting topological phases. 
	\end{abstract}
	\date{\today}
	
	\maketitle
Spin chirality, defined as a three-spin mixed vector product $\vec S_1\cdot (\vec S_2\times\vec S_3)$, is a key quantity 
for various order types in solids. 
Chirality arises in 
staggered spin phases resulting from spontaneous breaking of spin rotation symmetry, as well as in phases displaying 
breaking of time reversal symmetry without breaking the spin rotation symmetry [see, e.g., Refs.
\cite{Wen1989,Shraiman1989,Sen1995,Martin2008,Sedrakyan2015}]. When present, chirality 
leads to interesting properties of the ordered state. Notably, it generates a Berry phase for electrons moving through the texture and coupled to it by spin exchange interactions. 
The geometric gauge fields and emergent  electromagnetism arising in this way are manifested through anomalous Hall effect\cite{Ye1999,Nagaosa2012b,Nagaosa2012,Chen2014,Ishizuka2018}. In magnetic dielectrics, chiral interaction stabilizes frustrated and noncollinear spin phases, as well as a rich variety of  spin liquid phases\cite{Lee2005,Motrunich2006,Bulaevskii2008,Katsura2010,Lai2010}. In spin liquids, geometric magnetic fields resulting from spin chirality are predicted to drive quantum oscillations of spinons
\cite{Sodemann2018,He2023}. 
The intriguing physics of spin chirality raises the question of what general conditions allow it to emerge in a solid.

\begin{figure}
		\centering
		\includegraphics[width=0.9\linewidth]{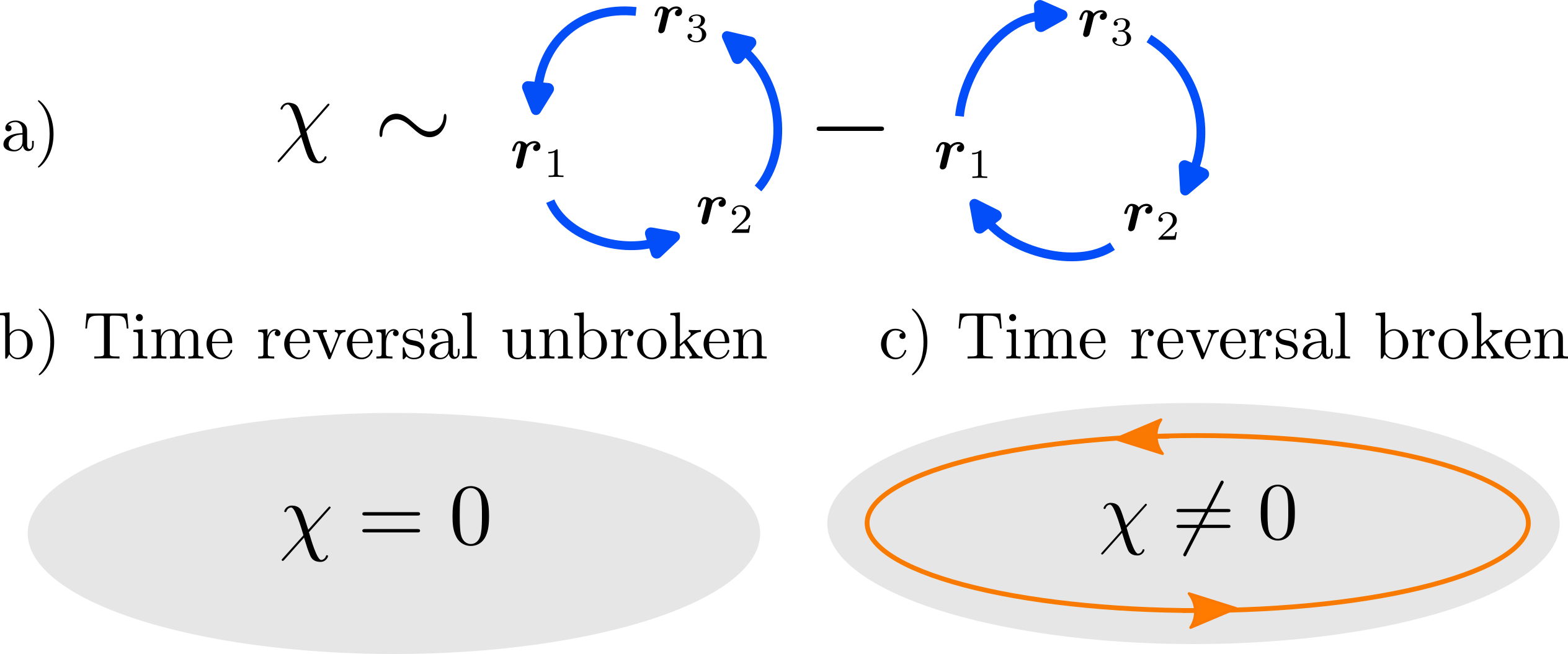}
		\caption{
		a) Spin chirality, Eq.\eqref{eq:chiral_effect}, in spin-unpolarized metals 
		arising in a state with broken time reversal symmetry (TRS). Chirality originates from stirring dynamics due to 
		circulating orbital 
		currents, which  
		drive three-particle chiral exchange processes [see Eq.\eqref{eq:chirality_permutations}]. Shown are diagrammatic contributions given in 
		 Eq.\eqref{eq:GGG-GGG}.  b), c) In systems where TRS breaking occurs 
		through a phase transition, such as, e.g., topological superconductors or valley-polarized phases in graphene multilayers, spin chirality emerges simultaneously with orbital magnetism even if the ordered state remains spin-unpolarized.
		}
		\label{fig:1}
		\vspace{-3mm}
	\end{figure}

The examples above might suggest that chirality requires fine-tuned prerequisites, such as spin-dependent interactions, specific spin order or crystal lattice symmetry. However, as we will demonstrate, chirality is a much more general property than these examples indicate. We will show that nonzero chirality robustly arises in any system with broken time reversal 
symmetry (TRS), without needing any other special properties, as illustrated in Fig.\ref{fig:1}. Indeed, from a symmetry point of view, chirality is a pseudoscalar under spin rotations. Therefore, it is expected to be nonzero even if the Hamiltonian is spin-independent and involves only orbital degrees of freedom, even if the ground state $\left. |0\ra$ is spin-unpolarized and invariant under spin rotations. 

This argument predicts that metals with broken TRS must exhibit nonzero three-point spin chirality\cite{foornote1}:
\be\label{eq:chiral_effect}
\chi(\vec r_1,\vec r_2,\vec r_3)=
\la 0|\,\vec s(\vec r_1) \cdot [\vec s(\vec r_2) \times \vec s(\vec r_3)]\,|0\ra\ne0
,
\ee 
where $\vec s(\vec r) =\sum_{\alpha\beta}\bar\psi_{\alpha}(\vec r)\vec \sigma_{\alpha\beta}\psi_{\beta}(\vec r)$ is electron spin density, defined, for conciseness, without the usual $\hbar/2$ factor. 
Below, this conclusion will be justified microscopically for generic spin-unpolarized metals. 
To emphasize the generality of chirality beyond specific examples, we consider three cases: (i) Berry phase in coordinate space due to 
magnetic field, geometric or electromagnetic, which couples to orbital degrees of freedom but not to electron spin; (ii) Berry phase due to Berry curvature in $k$-space of a conduction band; (iii) Berry phase due to the superconducting order parameter in topological superconductors with spin-unpolarized ($S=0$) pairing order. In all three cases, the Fermi sea is invariant under SU(2) spin rotations, i.e., has symmetry of a spin singlet. Spin-dependent interactions present in realistic systems, if weak, will leave nonzero chirality unaffected.

Physically, this effect 
represents a reciprocal of the familiar effect of geometric phases 
of chiral spin textutres driving orbital currents\cite{Ye1999,Nagaosa2012b,Nagaosa2012,Chen2014}.
Namely, orbital dynamics in a ground state with broken TRS 
(loosely described as stirring) 
leads to chiral spin correlations. 
Intuitively, this behavior can be understood with the help of an identity\cite{Wen1989} that links spin chirality operator to three-spin permutations: 
\be\label{eq:chirality_permutations}
H^{(3)}=\vec s_1\cdot[\vec s_2\times\vec s_3] 
=2iP_{123}-2iP_{321}
,
\ee
which is valid for any three spin-$1/2$ degrees of freedom, with  $P_{jkl}$ being cyclic permutations. 
This result, 
familiar 
for localized spins\cite{Wen1989}, here is applied to orbital dynamics of  carriers. When read right to left, it implies that stirring induces chiral three-spin exchange and enables 
spin chirality. This is indeed what emerges from the analysis. 

As we will see, the three-point spin chirality given in Eq.\eqref{eq:chiral_effect} 
can be expressed through electron Green's functions in coordinate space,
\be\label{eq:Greens_function}
G(\epsilon,\vec r_j,\vec r_k)=
\la \vec r_k \left|\frac1{i\epsilon-H}\right|\vec r_j\ra
,
\ee
where $H$ is the carrier Hamiltonian that depends on orbital degrees of freedom but not spin.
Diagrammatically, this quantity
is given by a difference of contributions due to particles propagating between points $\vec r_1$,  $\vec r_2$,  $\vec r_3$ clockwise and counterclockwise: 
\be\label{eq:GGG-GGG}
\chi(\vec r_1,\vec r_2,\vec r_3) \! = \! 
2i
T\sum_\epsilon \!\!
 \tr G_{12}G_{23}G_{31}\!-\!\tr G_{21}G_{13}G_{32}
,
\ee
where $G_{jk}$ is a shorthand for the Green's function in Eq.\eqref{eq:Greens_function} and $\tr$ 
denotes trace over band indices of $H$. 

As a local property sensitive to TRS, spin chirality can be used to detect local TRS breaking. Usually, TRS breaking in metals is detected by the Hall effect or optical Kerr effect, or by torque magnetometry (e.g., see \cite{Modic2018, Mumford2020, Saykin2023} and references therein). However, in many systems of interest, the 
ordered states feature intertwining domains of opposite polarization that are challenging to detect macroscopically. In such cases, local probes have a distinct advantage, as demonstrated, 
e.g., in the recent detection of the Chern mosaic and valley polarization in 
moir\'e graphene\cite{Grover2022}. Below we consider magnetic adatoms/impurities at the surface of a 2D electron system, coupled to 
electron spins by an s-d Hamiltonian. We find that spin chirality in an electron gas, Eq.\eqref{eq:chiral_effect}, governs chiral exchange interactions between adatom spins. 

			\begin{figure}
		\centering
		\includegraphics[width=0.90\linewidth]{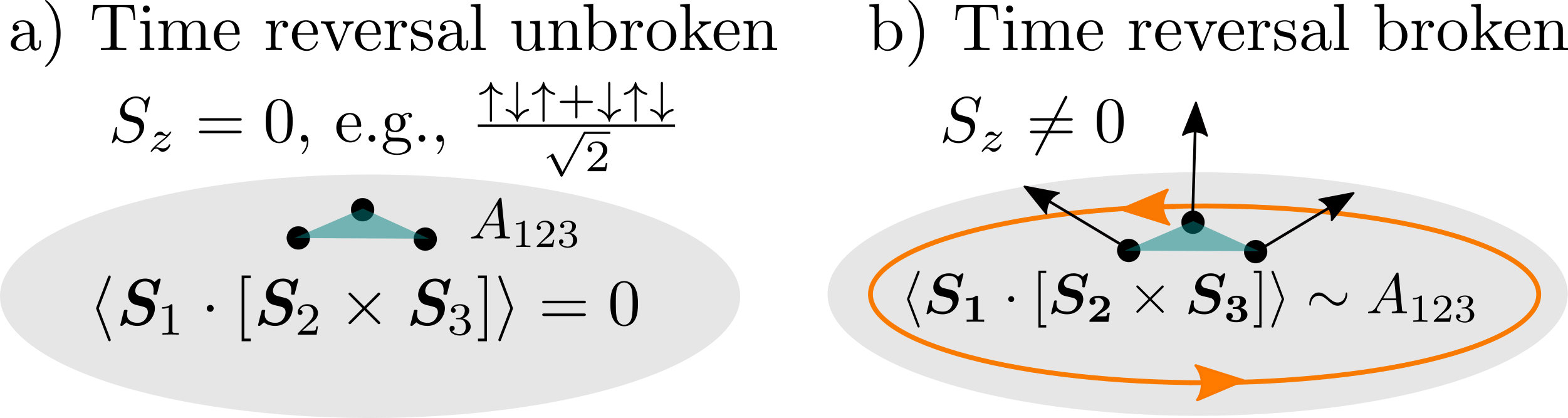}
		\caption{
		Trion of magnetic adatoms as a chirality meter. The three-spin Hamiltonian of a magnetic trion, Eq.\eqref{eq:spin_trion}, features chiral interaction $\lambda \vec S_1\cdot\lp \vec S_2\times \vec S_3\rp$ of strength $\lambda$ proportional to local chirality value. In the presence of a DM interaction, Eq.\eqref{eq:H_DM}, the trion develops a nonzero net magnetization of a sign proportional to $\lambda D$, Eq.\eqref{eq:trion_sz}. 
		}
		\label{fig:2}
		\vspace{-3mm}
	\end{figure}


Specifically, for trions composed of three adatoms, we obtain a three-spin analog of the RKKY interaction:
\be\label{eq:spin_trion}
H=\sum_{i\ne k}J_{ik}\vec s_i\cdot \vec s_k 
+\lambda \vec s_1\cdot(\vec s_2\times \vec s_3)+H_{\rm DM}
\ee
where $J_{ik}>0$ are pairwise RKKY interactions and the chiral coupling $\lambda$ is equal, up to s-d exchange coupling, to the three-point spin chirality in Eq.\eqref{eq:GGG-GGG}. The last term is the Dzyaloshinsky-Moriya (DM) interaction 
\be\label{eq:H_DM}
H_{\rm DM}=D(\vec s_1\times \vec s_2)^z+D(\vec s_2\times \vec s_3)^z+D(\vec s_3\times \vec s_1)^z
\ee
generally required at a 2D interface by symmetry\cite{Dzyaloshinsky1957,Moriya1960,Kang_Lee2024}.

As illustrated in  Fig.\ref{fig:2}, the chiral three-spin interaction in Eq.\eqref{eq:spin_trion} is directly sensitive to local chirality values, 
resulting in noncollinear three-spin ground states with a nonzero $\la s^z\ra$ magnetization. 
Indeed, in the presence of a DM coupling but in the absence of the chiral term, $\lambda=0$, the Hamiltonian obeys TRS and features a double-degenerate spin-$1/2$ ground state. This doublet is split on introduction of the chiral term, resulting in TRS breaking and a finite magnetization 
at $T=0$ is $\la s^z\ra=
{\rm sgn}\lambda$, of a sign governed by 
$\lambda$. Magnetization gradually decreases as temperature grows. In a realistic case of slightly unequal couplings $J_{ik}$ and for $D,\,\lambda\ll J_{ik}$, Eq.\eqref{eq:spin_trion} predicts a decreasing $T$ dependence \cite{footnote2}  
$\la s^z\ra=
\tanh( \lambda D/\tilde J k_BT)$, 
with $\tilde J$ given by, up to a constant of order one, the variance of $J_{ik}$. 
After restoring units of $\hbar/2$ for spin-$1/2$ angular momentum, this translates into magnetization
\be \label{eq:trion_sz}
m_z=\frac{g\mu_B}2 \tanh( \lambda D/\tilde J k_BT)
.
\ee
This result 
holds in a wide  temperature interval $0\le k_BT\lesssim \tilde J$.
Magnetization of adatoms induced by chirality is an effect that can be measured by state-of-the-art spin-sensitive STM probes 
\cite{Manassen1989,Madhavan1998,Wiesendanger2009,Gonzalez-Herrero_2016,Baumann2015}. 



General properties of chirality, as given in Eq.\eqref{eq:GGG-GGG}, 
can be illustrated in the context of Berry phase in coordinate space due to a magnetic field, 
whether geometric or electromagnetic, coupled to orbital degrees of freedom but not to spin. As a simplest example, we consider carriers in a parabolic band, $H=\frac1{2m}(\vec p-\vec a)^2$. However, as will become clear, the results are considerably more general. 

At a weak magnetic field $\vec b=\nabla\times \vec a$, $\vec b\parallel \hat {\vec z}$, the Green's function is given by a product of the Green's function for $b=0$ and an Aharonov-Bohm phase factor evaluated for a straight path connecting the end points\cite{AGD}:
\be
G(\vec r_j,\vec r_k)=e^{i\theta_{jk}}
G^{0}(|\vec r_j-\vec r_k|)
,\quad
\theta_{jk}=\int_{\vec r_j}^{\vec r_k} \vec a\cdot d\vec r
,
\ee
where the dependence of $G$ and $G^0$ on  $\epsilon$
was suppressed for conciseness 
(we set $\hbar=1$, restoring dimensional units later). This quasiclassical expression, sometimes referred to as ``eikonal model'', is applicable to both nonquantizing and quantizing fields, so long as the distance between the end points is greater than the Fermi wavelength\cite{AGD}. 
The two 
contributions in Eq.\eqref{eq:GGG-GGG} then acquire opposite phases $\theta_{123}=\theta_{12}+\theta_{23}+\theta_{32}$ and $\theta_{213}=-\theta_{123}$. 
This gives chirality proportional to $ie^{i\theta_{123}}-ie^{i\theta_{213}}=-2\sin \theta_{123}$:
\begin{align}\label{eq:GGG-GGG_phases}
&\chi(\vec r_1,\vec r_2,\vec r_3) =-4 \sin \theta_{123}
T\sum_\epsilon G^{0}_{12}G^{0}_{23}G^{0}_{31}
\nonumber \\ 
&=-4 \sin \lp 2\pi \frac{bA_{123}}{\phi_0}\rp
T\sum_\epsilon G^{0}_{12}G^{0}_{23}G^{0}_{31}
,
\end{align}
where 
the Aharonov-Bohm phase gained on going around the triangle  $(\vec r_1, \vec r_2, \vec r_3)$ was expressed through the triangle area  $A_{123}$ and the flux quantum $\phi_0$. 
This result, which links chirality to the flux of $b$ through the triangle $(\vec r_1, \vec r_2, \vec r_3)$, applies to electrons in an external magnetic field regardless of its origin. E.g., it describes electrons in a geometric field $b$ due to 
isospin textures such as those discussed recently for twisted TMD homobilayers \cite{Wu2019,Morales2024} and isospin-polarized Bernal bilayers\cite{Dong2022}. 

The result in Eq.\eqref{eq:GGG-GGG_phases} clarifies the geometric meaning of chirality and its general properties. Chirality changes sign when the triangle orientation is reversed; it also vanishes when time reversal symmetry is restored  at $b=0$. When $b$ is finite but small, or when the three points are close to each other, chirality is proportional to the triangle area $A_{123}$. 
It is a local quantity, taking large values at distances $|\vec r_j-\vec r_k|$ comparable to the Fermi wavelength, 
and falling off quickly at larger distances. E.g., for a parabolic band, 
$G^0(\epsilon,\vec r_j,\vec r_k)\sim -\frac{2m}{\hbar^2}\log|\vec r_j-\vec r_k|/\lambda_F$ at short distances, decaying rapidly at distances greater than $\lambda_F$.

Vorticity of orbital currents, the effect  
responsible for spin chirality, manifests through the angular dependence of 
$G(\epsilon,\vec r_j,\vec r_k)$ allowing particles to gain phase on going around a loop. In the example above, the phase was due to 
a magnetic field, positive or negative depending on the loop orientation. 
Similar behavior will also occur for carriers in bands equipped with $k$-space Berry curvature, in the absence of magnetic field. Here, we demonstrate it for the Haldane model\cite{Haldane1988}, 
representing a gapped graphene bandstructure 
created by a staggered magnetic flux pattern. 
Haldane model 
captures the 
behavior in TRS-broken phases of valley-polarized  graphene 
and many other systems [see \cite{Grover2022} and references therein].


The Hamiltonian for this model is of the form $H(\vec p)=\vec d(\vec p)\cdot \vec \tau$, where $\vec \tau=(\tau_1,\tau_2,\tau_3)$ are isospin $2\times 2$ Pauli matrices, and $\vec d(\vec p)$ is a vector that, as $\vec p$ varies in the Brillouin zone, sweeps a $4\pi$ solid angle.  It is instructive to focus on the small-gap regime, achieved by tuning the staggered flux 
in the model, where the Berry curvature distribution in $k$-space is concentrated near points $K$ and $K'$. In this case, since relevant electron states have low energies with momenta close to $K$ and $K'$, the full graphene band can be reduced to a gapped Dirac Hamiltonian, separately for each valley:
	\be\label{eq:H_monolayer}
	H_\eta(\vec p) = \lp \begin{array}{cc}
		m_\eta & vp_{-}\\
		vp_{+} & -m_\eta
	\end{array} \rp 
	,\quad p_\pm=\eta p_x\pm i p_y
	,
	\ee
where $\eta=\pm 1$ for valleys $K$ and $K'$, respectively, 
$p_{x,y}=-i\partial_{x,y}$, and for simplicity we work with half-filled band, $\mu=0$. The quantities $m_\eta$, which are functions of perturbations breaking P and T symmetries, in general have different values and different signs. 
Berry curvature, which is nonzero in each valley, generates a nonzero net orbital magnetization unless $m_{\eta=+1}$ and $m_{\eta=-1}$ are equal in magnitude and sign. 


The angular dependence of the two-point Green's functions, Eq.\eqref{eq:Greens_function}, 
is obtained by a standard algebra: 
\begin{align}
&G(\epsilon,\vec r)=\int \frac{d^2p}{(2\pi)^2}
\frac{e^{i\vec p\vec r}}{i\epsilon-m_\eta \tau_3-\eta p_1\tau_1-p_2\tau_2}
\nonumber \\
& =\lp -i\epsilon-m_\eta \tau_3+\tau_1 i\eta\p_x+\tau_2 i\p_y\rp \frac1{2\pi} K_0(r /\xi)
,
\end{align}
where $\xi=(\epsilon^2+m_\eta^2)^{-1/2}$ and $K_0(z)$ is the modified Bessel function of the second kind 
(for conciseness, we set $v=1$).
Using the identity $\frac{d}{dz}K_0(z)=-K_1(z)$ to evaluate derivatives, and defining scaled distance $z=r/\xi$, gives
\be
G(\epsilon,\vec r)=-\frac{ i\epsilon+m_\eta \tau_3}{2\pi} K_0(z)
- \frac{\tau_1 i\eta x+\tau_2 iy 
}{2\pi \xi^2} \frac{ K_1(z)}{z}
.
\ee
This Green's function is a $2\times2$ matrix of a general form 
$G(\epsilon,\vec r)=A_{\vec r}\tau_0+B_{\vec r}\tau_3+C_{\vec r}\tau^++C^*_{\vec r}\tau^-$. Because of the angular $\vec r$-dependence, such matrices do not commute for noncollinear vectors $\vec r$. 
With this matrix structure, the chirality in Eq.\eqref{eq:GGG-GGG} readily takes the form of a determinant 
\begin{align}\nonumber
&\chi(\vec r_1,\vec r_2,\vec r_3) =4i T\sum_\epsilon {\rm det}\lp \begin{array}{ccc}
B_{\vec r_1-\vec r_2} & B_{\vec r_2-\vec r_3}& B_{\vec r_3-\vec r_1}
\\
C_{\vec r_1-\vec r_2} & C_{\vec r_2-\vec r_3}& C_{\vec r_3-\vec r_1}
\\
C^*_{\vec r_1-\vec r_2} & C^*_{\vec r_2-\vec r_3}& C^*_{\vec r_3-\vec r_1}
\end{array}\rp
\\ \label{eq:chi=det}
& 
= -T\sum_{\eta,\epsilon}
\frac{ A_{123}}{\pi^3\xi^4} m_\eta \eta \lb K_0(z_{12})\tilde K_1(z_{23})\tilde K_1(z_{31})
\right.
\\ \nonumber
& \left.+\tilde K_1(z_{12})K_0(z_{23})\tilde K_1(z_{31})
+\tilde K_1(z_{12})\tilde K_1(z_{23}) K_0(z_{31}) \rb
,
\end{align}
where 
$\tilde K_1(z) 
\equiv K_1(z)/z$ and $z_{jk}$ denotes scaled distances $|\vec r_j-\vec r_k|/\xi$.
This quantity is proportional to $A_{123}$, the area of the oriented triangle $(\vec r_1,\vec r_2,\vec r_3)$ 
(see Fig.\ref{fig:2}) 
and is dominated by the valley for which $m_\eta \eta$ is greatest, 
as expected. 
The chirality sign is determined
solely by the sign of $A_{123}m_\eta \eta$.
The behavior at short distances can be inferred from 
the asymptotic expansion $K_0(y)=-\ln \frac{y}2-\gamma+O(y^2)$. Evaluating derivatives 
yields a coordinate dependence 
$\chi(\vec r_1,\vec r_2,\vec r_3)$ that is large at short distances and falls off quickly as distances grow. 

Next, we outline a 
general microscopic framework linking spin chirality to Green's functions. We will use 
a generating functional 
\be\label{eq:Z_generating}
Z[\vec h(\vec r)]={\rm Tr} \lb {\rm Texp}\lp -\int_0^\beta \mathcal{H}(t)dt\rp\rb
,
\ee
where the Hamiltonian includes a source term coupled to the spin density operator
$\vec s(x) =\sum_{\alpha\beta}\bar\psi_{\alpha}(x)\vec \sigma_{\alpha\beta}\psi_{\beta}(x)$: 
	\be\label{eq:H_generating}
	\mathcal{H}(t) = 
	\int \!\! d^2r \bar\psi(x) \lb H(\vec p)\otimes 1_{\rm S} +\vec h(x)\cdot \vec \sigma 
	\rb \psi(x)
	,
	\ee	
where $x=(\vec r,t)$ labels space and time coordinates, and $\vec \sigma=(\sigma_1,\sigma_2,\sigma_3)$ are Pauli matrices. An  identity matrix in spin variables $1_{\rm S}$ is introduced to emphasize that 
$H(p)$ does not depend on electron spin. 
For conciseness, this quantity is suppressed below. 
Likewise, when carriers populate several valleys, as in Eq.\eqref{eq:H_monolayer}, we assume that they couple to a single source field $\vec h(x)$. Generalization to valley-specific $\vec h(x)$ is straightforward.

Spin correlators can be found by Taylor expanding the log of the generating functional \cite{AGD,Coleman2015}:
\begin{align}
F[\vec h(x)]&=-\log \frac{Z[\vec h(x)]}{Z[0]}=\sum_{n>0}\frac1{n}\int d^3x_1...d^3x_n
\\ \nonumber
& \times h^{i_1}(x_1)...h^{i_n}(x_n)\la\la s^{i_1}(x_1)...s^{i_n}(x_n)\ra\ra
,
\end{align}
where  
$\la\la\ldots\ra\ra$ denotes irreducible correlators, or cumulants. Of special interest for us will be the third-order term since it is this term that gives the chirality density. 

In our problem, the generating functional can be analyzed most easily using fermionic field representation 
\be
Z=\int \mathcal{D}[\psi,\bar\psi] e^{-\int dt d^2r \bar\psi(x) \lb \p_t+H(\vec p) +\vec h(x)\cdot \vec \sigma\rb \psi(x) }
,
\ee
where the time integral is taken over $0<t<\beta$. 
After taking the Gaussian integral over the fermion variables, the 
quantity $Z$ is expressed as a fermion determinant:
\be\label{eq:fermionic_determinant}
Z[\vec h(x)]={\rm det}\lb \p_t+H(\vec p) +\vec h(x) \cdot \vec \sigma 	
\rb
.
\ee
This transformation simplifies the analysis of the generating functional in terms of  fermion contributions.

The generating functional given as a fermionic determinant, Eq.\eqref{eq:fermionic_determinant},
yields an expansion:
\begin{align}\label{eq:h_expansion_general}
& F[\vec h(x)]=-\log \frac{Z[\vec h(x)]}{Z[0]}=\sum_{n>0}\frac1{n}\int d^3x_1...d^3x_n 
\\ \nonumber
& {\rm tr}\lb (\vec h(x_1)\cdot\vec \sigma) G(x_1-x_2)...(\vec h(x_n)\cdot\vec \sigma) G(x_n-x_1)\rb
,
\\ \nonumber
& G(x-x')=T\sum_\epsilon \int \frac{d^2p}{(2\pi)^2}\frac{e^{i\vec p(\vec r-\vec r')-i\epsilon(t-t')}}{i\epsilon-H(\vec p)}
,
\end{align}
where $\epsilon$ is a discrete frequency. The Green's functions are scalars for single-band problems and 
matrices with band indices for matrix Hamiltonians $H(\vec p)$. 
The trace is taken over both spin and band variables. 

To analyze the power series expansion in Eq.\eqref{eq:h_expansion_general}, we introduce Fourier harmonics of the source field:
\be\label{eq:h_wk}
h^{i}(x)=\sum_{\omega,\vec k} h^{i}_{\omega,\vec k}e^{-i\omega t+i\vec k\cdot\vec r}
.
\ee
Below, for simplicity, we focus on time-independent source fields $\vec h(x)$, i.e., we retain only the $\omega=0$ harmonics 
in Eq.\eqref{eq:h_wk}. This limit describes static equilibrium correlation functions of spin densities in the metal. Dynamical correlations will be discussed elsewhere. 

In this case, the generating functional is 
expressed through products of Green's functions which are all taken at the same $\epsilon$ and integrated over $\epsilon$:
\begin{align}
F[\vec h(x)]&=
\sum_{n>0}\frac1{n}\int d^2r_1...d^2r_n T\sum_\epsilon 
{\rm tr}\lb (\vec h(\vec r_1)\cdot\vec \sigma)\right.
\nonumber \\
\label{eq:F[h]}
& \left. \times G(\epsilon,\vec r_1-\vec r_2)...(\vec h(\vec r_n)\cdot\vec \sigma) G(\epsilon,\vec r_n-\vec r_1)\rb
,
\nonumber \\ 
& G(\epsilon,\vec r-\vec r')=\int \frac{d^2p}{(2\pi)^2}\frac{e^{i\vec p(\vec r-\vec r')}}{i\epsilon-H(\vec p)}
.
\end{align}
These expressions can be simplified by evaluating trace over spin variables as
\be
{\rm tr\,} \sigma^j=0,
\ \ 
{\rm tr\,} \sigma^{j}\sigma^{k}=2\delta_{jk},
\ \ 
{\rm tr\,} \sigma^{j}\sigma^{k}\sigma^{l}=2i\varepsilon_{jkl},
\ee
where $\varepsilon_{jkl}$ is the antisymmetric tensor. 
Therefore, the $n=1$ contribution vanishes, the $n=2$ contribution gives the standard RKKY pair spin correlator. The $n=3$ contribution, which encodes chirality, takes the form:
\begin{align}\label{eq:delta_F3_r}
\delta F_3&=\int d^2rd^2r'd^2r''
\frac{2i}3 (\vec h_{\vec r}\times \vec h_{\vec r'})\cdot \vec h_{\vec r''}
\\ \nonumber
& \times T\sum_\epsilon  
{\rm tr}\, G(\epsilon,\vec r-\vec r')G(\epsilon,\vec r'-\vec r'')
G(\epsilon,\vec r''-\vec r)
.
\end{align}
The mixed vector product structure of 
this expression clearly indicates a relationship with chirality. 

This result 
yields a closed-form expression for a three-point spin correlator $\la s^j(\vec r_1)s^k(\vec r_2)s^l(\vec r_3)\ra$ 
after identifying coordinates  in the integral $(\vec r,\vec r',\vec r'')$ with  points $(\vec r_1,\vec r_2,\vec r_3)$. Altogether, there are six contributions, three of one sign and three of the opposite sign: 
\begin{align}\label{eq:<sss>}
&\la s_j(\vec r_1)s_k(\vec r_2)s_l(\vec r_3)\ra=\frac{2i}3 \varepsilon_{jkl}
T\sum_\epsilon 
\\ \nonumber
&\times 
\sum_P (-1)^P {\rm tr}\, G(\epsilon,\vec r-\vec r')G(\epsilon,\vec r'-\vec r'')
G(\epsilon,\vec r''-\vec r)
\end{align}
where the sum $\sum_P$ is taken over six permutations $(\vec r,\vec r',\vec r'')=P(\vec r_1,\vec r_2,\vec r_3)$, with the sign $(-1)^P$ plus and minus for permutations of a positive and negative parity, respectively.  
This provides microscopic derivation of the result for the three-point spin chirality as given in Eq.\eqref{eq:GGG-GGG}. 


The diagrammatic approach used to derived Eq.\eqref{eq:GGG-GGG} also describes the 
chiral RKKY interaction between magnetic adatoms mediated by electrons, 
as given in  Eq.\eqref{eq:spin_trion}. 
Indeed, 
spins $\vec S_i$ localized at $\vec r=\vec r_j$ and coupled to electron spins by the s-d Hamiltonian\cite{Hewson2008}
\be
{\cal H}=\sum_p \psi^\dagger(\vec p)H(\vec p)\psi(\vec p)+J\sum_{\vec r_j}\vec S_j \vec s(\vec r_j)
,
\ee
is identical in form to the generating functional problem 
in Eqs. \eqref{eq:Z_generating} and \eqref{eq:H_generating}, 
as verified by identifying the source field $\vec h(x)$ with $\sum_j J \vec S_j\delta(\vec r-\vec r_j)$. 
Integrating out fermion degrees of freedom yields an effective interaction Hamiltonian for localized spins. At leading order in $J$, ignoring higher-order Kondo-like contributions, 
we obtain a sum of terms that match cumulants found from the generating functional. 
This term-by-term identification is justified at small $J$, when spin dynamics governed by $J$ is slow on the characteristic time scales of electrons, $t\sim |\vec r_j-\vec r_k|/v_F$. 
This predicts the 
familiar two-spin RKKY interaction and the chiral three-spin coupling $\lambda=J^3\chi(\vec r_1,\vec r_2,\vec r_3)$ in Eq.\eqref{eq:spin_trion}.


Next, we discuss chiral spin effects induced by topological superconductivity in metals for pairing with nonzero orbital angular momentum.
In this case, even if TRS is unbroken in the normal state, it is broken in the superconducting phase as a result 
of the order parameter phase winding around the Fermi surface.
As is well known, this translates into a nonzero $k$-space Berry curvature 
for the BdG Hamiltonian of the system. 
Broken TRS results in a chiral spin effect present even when the superconducting order is spin-singlet and invariant under SU(2) spin rotations.
Here, in complete analogy with the discussion of the Dirac band above, spin chirality is enabled by TRS being broken for orbital degrees of freedom, in a complete absence of spin-orbit interactions. 

Further illustration of the generality of the link between Berry phase and spin chirality,
and a system displaying interesting physics, is provided by topological superconductivity. As a simple example, we consider pairing in a parabolic band with nonzero angular momentum, 
\be\nonumber
H=\sum_{\vec p,\sigma}\epsilon(\vec p)c^\dagger_{\vec p,\sigma}c_{\vec p,\sigma}
-\sum_{\vec p,\vec p'}V(\vec p-\vec p')c^\dagger_{-\vec p'\downarrow} c^\dagger_{\vec p'\uparrow} c_{\vec p\uparrow} c_{-\vec p\downarrow}
,
\ee
where $\sigma=\uparrow,\downarrow$. The angle-dependent pairing interaction $V(\vec p-\vec p')$ is such that it favors exotic superconductivity with $L\ne0$ and $S=0$. The paired state is described by a $4\times 4$ Bogoliubov-de Gennes (BdG) Hamiltonian
\be\label{eq:H_BdG}
H_{\rm BdG}(\vec p)=\lp \begin{array}{cc}\xi_{\vec p}& \Delta_{\vec p}\sigma_y\\ \Delta^\ast_{\vec p}\sigma_y & -\xi_{\vec p}
\end{array}\rp
,\quad
\xi_{\vec p}=\epsilon_{\vec p}-\mu
,
\ee
where $\sigma_y$ describes spin-singlet pairing and 
the order parameter 
$\Delta_{\vec p}$ 
phase winds around the Fermi surface,
\be
\Delta_{\vec p}=\Delta'_{\vec p}+i\Delta''_{\vec p}
=|\Delta_{\vec p}| e^{i\theta_{\vec p}},\quad
\oint d\theta_{\vec p}=2\pi L
.
\ee 
Since this Hamiltonian features nonzero Berry curvature, by the symmetry argument given above and the unbroken SU(2) symmetry notwithstanding, it is expected to give nonzero spin chirality. 

The analysis can be carried out in the same way as for
Haldane model. We use the generating functional as given in
Eqs. \eqref{eq:Z_generating} and \eqref{eq:H_generating}, 
 with the BdG Hamiltonian 
 and coupling of the source field to electron spin density written as in a superconductor with magnetic impurities\cite{Abrikosov1961} 
 \begin{align} 
 & H=H(\vec p)+\vec h(r)\cdot\lp \frac{1+\tau_3}2\vec s(\vec r)+\frac{1-\tau_3}2
 \sigma_y \vec s(\vec r)\sigma_y\rp
 ,\quad
 \nonumber \\ 
 &
 \vec h(\vec r)=\sum_j J \vec S_j\delta(\vec r-\vec r_j).
 \end{align} 
This spin structure accounts for the fact that spin states in the BdG Hamiltonian are $(\psi_1,\psi_2)$ and $(\psi_1,-\psi_2)$, for particles and holes, respectively \cite{Yu1965,Shiba1968,Rusinov1969}. This Hamiltponian, after integrating out fermions, 
generates a result similar to Eq.\ref{eq:GGG-GGG}, 
where $G(\epsilon,\vec r-\vec r')$ are 
matrix Green's functions of the superconductor. The general expression 
\begin{align}
&G(\epsilon,\vec r-\vec r')=
\int\frac{d^2p}{(2\pi)^2}\frac{e^{i\vec p\cdot(\vec r-\vec r')}}{i\epsilon
-H_{\rm BdG}(\vec p) 
}
\\ \nonumber
&
=A_{\epsilon,\vec r-\vec r'}\tau_0+B_{\epsilon,\vec r-\vec r'}\tau_3+C_{\epsilon,\vec r-\vec r'}\tau^{+}\sigma_y+C_{\epsilon,\vec r-\vec r'}^*\tau^{-}\sigma_y
\end{align}
simplifies at $p_F|\vec r-\vec r'|\gg 1$ and $\Delta\ll E_F$. In this case, 
\be\nonumber 
\lp\begin{array}{c} A_{\epsilon,\vec r} \\ B_{\epsilon,\vec r} \\ C_{\epsilon,\vec r}\end{array}\rp=\sqrt{\frac{p_F}{2\pi |\vec r|}}\frac{e^{-|\vec r|/\xi(\epsilon)}}{v_F^2 }
\lp\begin{array}{c} -i\epsilon\xi(\epsilon) \cos \phi_{\vec r} 
\\ v_F\sin \phi_{\vec r} 
\\ -\Delta\xi(\epsilon) \frac{(x+iy)^2}{|\vec r|^2}\cos \phi_{\vec r} 
\end{array}\rp
\ee
where $\xi(\epsilon)=v_F/\sqrt{\epsilon^2+\Delta^2}$, $\tau^{\pm}=(\tau_x\pm i\tau_y)/2$ and $\phi_{\vec r}=p_F|\vec r|-\frac{\pi}4$. For concreteness, we focus on topological ${\rm d+id}$ pairing, $L=2$. 

To obtain spin chirality, we employ the determinant formula used  for Haldane model, 
which gives \cite{footnote_sigma_y}
\be
\chi(\vec r_1,\vec r_2,\vec r_3)=4i T\sum_{\epsilon,P}(-)^P 
B_{\epsilon,\vec r-\vec r'} C_{\epsilon,\vec r'-\vec r''}C^*_{\epsilon,\vec r''-\vec r}
,
\ee 
with the sum taken over six permutations of three points 
$(\vec r,\vec r',\vec r'')=(\vec r_{P(1)},\vec r_{P(2)},\vec r_{P(3)})$. This predicts chirality proportional to the area of the oriented triangle $A_{123}$ times some functions of coordinates falling off at large $\vec r_j-\vec r_k$. 
This property follows from the identity
\be
(x-iy)^2(x'+iy')^2-{\rm c.c.} 
=4i(xy'-x'y)(xx'+yy')
.
\ee
Chirality oscillates with  $\lambda_F$ periodicity 
and falls off exponentially beyond the coherence length $\xi_0=v_F/\Delta$. 
Chirality values, after integration over $\epsilon$, are proportional to $|\Delta|/E_F$ times factors which are order-one at distances 
of the order $\lambda_F$, 
where chirality attains maximum value. 

In summary, spin chirality is a quantity expected on symmetry grounds whenever TRS is broken.
As we saw, 
orbital currents in a metal, normal or superconducting, present in the ground state as a result of TRS breaking, generate spin chirality even when the ground state is spin-unpolarized. 
The spin chirality effect can be probed using magnetic adatoms, and is particularly appealing as a probe of topological superconductivity. Presently, there exist many 
superconductors suspected to be topological, like UPt$_3$, $\text{Cu}_x\text{Bi}_2\text{Se}_3$, $\text{Sn}_{1-x}\text{In}_x\text{Te}$, moir\'e TMDs such as 2M-WS$_2$, twisted and stacked graphene layers, as well as cuprates \cite{Nandkishore2012-doped-graphene-chiral-SC, Lu2014-cuprate-topo-SC, Zhao2023-cuprate-topo-SC,Sato2017topo-super-review, Avers2020-cuprates-topo-SC, Li2021-2M-WSe2-topo-SC, Khosravian2024-moire-graphene-topo-SC}. Spin chirality may also act as a smoking gun probe to detect topology in superconducting materials like $\text{Sr}_2\text{RuO}_4$, where the topological nature of superconductivity is disputed \cite{NishiZaki1999, Mao2000, Yuguchi2002, Mackenzie2003, Sigrist2005, Hicks2014, Taniguchi2015, Mackenzie2017-SuRuO4-topo-SC-puzzle}. In such systems, while detecting TRS breaking by macroscopic measurements is challenging, local probes of spin chirality can serve as a reliable diagnostic of broken time reversal symmetry. 
 



%

We thank Leonid Glazman, Steven Kivelson, Patrick Lee, Andrew Millis, Chandra Varma, and Leo Radzikhovsky  for useful discussions. 
This work was supported by the Science and Technology Center for Integrated Quantum Materials, National Science Foundation grant No.\,DMR1231319 
and was performed in part at Aspen Center for Physics, which is supported by 
NSF grant PHY-2210452.


\begin{thebibliography}{9}
\bibitem{foornote1} More accurately, the three-point chirality is defined 
as a time-averaged quantity,
$\chi(\vec r_1,\vec r_2,\vec r_3)=\int dt_2dt_3\la 0|\vec s(\vec r_1,t_1)\cdot \lp \vec s(\vec r_2,t_2)\times \vec s(\vec r_3,t_3)\rp|0\ra$, integrated over times $t_2-t_1$ and $t_3-t_1$, with positions $\vec r_1$, $\vec r_2$ and $\vec r_3$ kept fixed. This matches the result of the microscopic analysis, as given in Eq.\eqref{eq:GGG-GGG}.
\bibitem{footnote2} A more accurate expression for magnetization induced by the DM and chiral interactions, obtained from the Hamiltonian in Eq.\eqref{eq:spin_trion}, is $\la s^z\ra=
\tanh\frac{\lambda D}{\tilde J k_BT}\tanh \frac{3\tilde J}{4k_BT}$. The first factor describes a sharply decreasing $T$ dependence, whereas the second factor is close to unity in the regime of interest  
($D,\,\lambda\ll J_{ik}$) and at not too high temperatures $T\ll \tilde J$.
\bibitem{footnote_sigma_y}
Here, it is important to consider the role of the quantity $\sigma_y$, which describes the spin-singlet structure of Cooper pairs 
as well as the relation between particle and hole spin wavefunctions. 
The dependence on $\sigma_y$ can be eliminated by transforming the hole part of the BdG spinor as $\psi_h'=\sigma_y\psi_h$. 
The transformed 
Hamiltonian does not involve $\sigma_y$,
\be
\tilde H(\vec p)=\vec d(\vec p)\cdot\vec \tau
=\lp \begin{array}{cc}\xi_{\vec p}& \Delta_{\vec p}\\ \Delta^\ast_{\vec p} & -\xi_{\vec p}
\end{array}\rp
.
\ee
This gives expressions identical to the determinant in Eq.\eqref{eq:chi=det} 
in terms of the transformed vector 
$\vec d(\vec p)$. 
Mapping to Bloch sphere defined by  vector $\vec d(\vec p)$ 
has a nonzero topological invariant $N=L$. This gives nonzero Berry curvature in $k$ space and nonzero spin chirality in position space.
\end{thebibliography}
\end{document}